\documentclass{PoS}
\let\ifpdf\relax

\usepackage{rotating}
\usepackage{mathrsfs}

\newcommand{\rt}{{\mathbf{r}_T}}
\newcommand{\xt}{{\mathbf{x}_T}}
\newcommand{\ut}{{\mathbf{u}_T}}

\newcommand{\vt}{{\mathbf{v}_T}}
\newcommand{\bt}{{\mathbf{b}_T}}
\newcommand{\yt}{{\mathbf{y}_T}}
\newcommand{\zt}{{\mathbf{z}_T}}
\newcommand{\ud}{\, \mathrm{d}}
\newcommand{\nc}{{N_\mathrm{c}}}
\newcommand{\tr}{\, \mathrm{Tr} \, }
\newcommand{\as}{\alpha_{\mathrm{s}}}

\title{The dynamics of strongly correlated gluons at high energies}

\ShortTitle{Dynamics of strongly correlated gluons at high energies}

\author{\speaker{Raju Venugopalan}\\ \thanks{Work supported by the Department of Energy under DOE Contract No.DE-AC02-98CH10886  and by an LDRD grant from Brookhaven Science Associates.}\\
        Physics Department, Brookhaven National Laboratory, Upton, NY 11973, USA\\
        E-mail: \email{raju@bnl.gov}}

\abstract{We describe some of the recent progress in our understanding of the dynamics of strongly correlated gluons at high parton densities. Computations in the Color Glass Condensate effective field theory provide a good description of inclusive and semi-inclusive final states in DIS, p+p and p+A collisions at small x. In nucleus-nucleus collisions, they provide an ab initio description of entropy generation, decoherence, isotropization and the onset of hydrodynamic flow. The successful description of a wide range of phenomena from RHIC to LHC is outlined and possibilities for more stringent tests noted.}

\FullConference{6th International Conference on Quarks and Nuclear Physics\\
		April 16-20, 2012\\
		Ecole Polytechnique, Palaiseau,  Paris}

\begin{document}

Why is it of fundamental importance to understand the properties of strongly correlated gluons at very high energies ? Simply, it is because it offers a promising route to understand empirically the fundamental structure of matter. Despite QCD being established as the fundamental theory of the strong interactions, we are still far from having clear pictures of hadrons and nuclei that transcend even limited contexts. For instance, the constituent quark model (and variants) gives quite a good picture of the spectrum of hadron resonances, but doesn't do as well with the internal spin dynamics of the proton, and has little of value to say about high energy scattering. The sophistication attained by lattice QCD is impressive to behold but there is scant hope that it too will have much to say about multiparticle production in the deep Minkowski region of the theory.

At the heart of the matter (no pun intended), are the outstanding problems of confinement and chiral symmetry breaking (and their interplay) that continue to thwart our developing a useful unified picture. In particular, there is an unresolved tension between the QCD parton model and the constituent quark picture. The former is not much use for spectroscopy but provides the framework for spectacularly successful computations of very high transverse momentum processes with all the ``unpleasant" stuff factorized into parton distributions.  These are universal and therefore provide predictive power. Perturbative computations however only reliably describe a small fraction of high energy cross-sections with the bulk parametrized by ``stringy" event generators. The latter, albeit extremely useful (in particular for searches beyond the standard model), at best provide limited insight into hadron structure. 

Genuine progress in understanding hadron structure can be achieved in the Regge-Gribov asymptotics of very high energies at fixed but large momentum resolution scales. In these asymptotics, a key feature of perturbative QCD is the rapid bremsstrahlung growth of gluon distributions. At fixed transverse resolution $Q^2$ of a probe, at a given parton momentum fraction $x$, this growth generates a phase space occupancy in the hadron approaching the maximal parametric value of $1/\alpha_S$ permitted by the stability of the theory. For each value of $Q^2$, there is a corresponding $x$ value where gluon distributions saturate for momenta below a saturation scale $Q_S (x)$ corresponding to maximal occupancies. This phenomenon called ``gluon saturation"~\cite{GLR} appears to be a fundamental feature of the theory\footnote{The phase space occupancy argument is specific to the light front parton model formulation of the theory. The arguments can be formulated as well for gauge invariant quantities such as field strengths squared or ``dipole" scattering amplitudes in the same framework.}.  In addition, the dynamical generation of a large scale $Q_S\gg \Lambda_{\rm QCD}$ suggests that the very non-perturbative properties of the hadron and hadronic scattering can be understood in weak coupling because $\alpha_S(Q_S) \ll 1$ in Regge-Gribov asymptotics. Furthermore, since the phase space occupancies are large ($1/\alpha_S(Q_S)\gg 1$), the saturation regime is a classical regime of the theory~\cite{MV}. 

Hadron and nuclear wavefunctions are many body configurations of (primarily) gluons in this high energy limit. Since we have a weak coupling theory, one can properly ask what the effective degrees of freedom are, what is the nature of their correlations, 
how precisely does the coupling run, and whether there is a universal fixed point common to hadrons and nuclei. Ultimately one would like to understand how this weak coupling yet non-perturbative dynamics at distance scales $1/Q_S$ is altered at larger distance scales by chiral symmetry and confinement. For a number of bulk observables, if $Q_S$ is large enough, can one anticipate that weak coupling computations can be performed with minimal contributions from these soft dynamics ?

The features mentioned as well as the kinematics of the high energy limit can be formulated in a weak coupling effective field theory (EFT) that describes the physics of saturated gluons as a Color Glass Condensate (CGC)~\cite{CGC-reviews}. The degrees of freedom\footnote{There is an increased understanding now of the dictionary between these d.o.f and those in the Reggeon Field 
Theory framework~\cite{Regge}.} in this EFT are dynamical gluon fields at small $x$ that couple coherently to static classical color charges at large $x$. Wilsonian renormalization group (RG) equations, derived from requiring that observables be independent of the separation in $x$ between sources and fields lead to an infinite hierarchy of evolution equations in $x$ for n-point Wilson line correlators averaged over dense color fields in the nucleus. For a physical observable defined by an average over all static color source configurations,
\begin{equation}
\langle {\cal O}\rangle_{_Y} \equiv \int [D\rho]\; W_{_Y}[\rho]\;
 {\cal O}[\rho] \; ,
\label{eq:CGC-exp}
\end{equation}
one can define a gauge invariant weight functional (``density matrix") of color sources $W_{_Y}[\rho]$ at rapidity $Y=\ln(x/x_0)\equiv \ln(x_0^-/x^-)$, where $x_0$ is the initial scale for small x evolution, satisfies the JIMWLK equation~\cite{JIMWLK} $\partial W_{_Y}[\rho] / \partial Y = {\cal H}\,W_{_Y}[\rho]$. From eq.~(\ref{eq:CGC-exp}) then, the energy evolution of the observable can be expressed as 
\begin{equation}
{\partial \langle {\cal O}\rangle_{_Y}\over \partial Y} = \langle {\cal H}\;
 {\cal O}\rangle_{_Y} \; .
\label{eq:CGC-BBGKY}
\end{equation}
The structure of  the JIMWLK Hamiltonian ${\cal H}$ is such that $\langle {\cal H}{\cal O}\rangle_{_Y}$ is an object distinct from $\langle {\cal O}\rangle_{_Y}$; one obtains an infinite hierarchy of evolution equations for operator expectation values $\langle{\cal O}\rangle_{_Y}$. Given appropriate initial conditions at large $x$, solutions of this Balitsky-JIMWLK hierarchy~\cite{Balitsky:1995ub,JIMWLK} allow one to compute a wide range of multi-particle final states in deeply inelastic scattering (DIS) and hadronic collisions.
A remarkable result is that the full dynamical content of this infinite hierarchy of correlation functions can be obtained explicitly  numerically by Langevin techniques~\cite{Blaizot:2002xy} as has been demonstrated in computations we shall describe shortly. 

 The simplest example of high energy scattering observables are the inclusive DIS structure functions 
$F_{2}$ and $F_{L}$ that are proportional to the forward scattering amplitude of a $q\bar{q}$  ``dipole'' on a nucleus. 
The forward dipole amplitude (dipole cross section) is given by 
\begin{eqnarray}
\sigma_\mathrm{dip.}(x, \rt) = 2\int \ud^2 \bt 
\times
\bigg< 1 
- \frac{1}{\nc}
\tr  V\left(\bt + \frac{\rt}{2}\right)
V^\dagger\left(\bt - \frac{\rt}{2}\right) \bigg> \, ,
\label{eq:dipole}
\end{eqnarray}
where $\rt = \xt -\yt$ is the transverse size of the dipole, 
$\bt = (\xt+\yt)/2$ is the impact parameter relative to the hadron and $V(\xt) = P\exp(ig \int dz^- \,{\rho(\xt,z^-)\over\nabla^2})$.  In general, the Balitsky-JIMWLK equation for the expectation value $S \equiv \langle \hat{S}\rangle$, with $\hat{S}(\xt-\yt) \equiv \frac{1}{\nc} \tr V(\xt) V^\dagger(\yt)$, also depends on the 
correlator $\langle \hat{S}(\xt-\zt)\hat{S}(\zt-\yt)\rangle$. In the large $\nc$ limit and for large nuclei, this factorizes to the form $S S$; numerical simulations of the B-JIMWLK hierarchy demonstrate this approximation to be much better than $1/\nc^2$ in all but a very small kinematic window~\cite{Rummukainen:2003ns,Dumitru:2011vk}. With this factorization, one obtains the closed form Balitsky-Kovchegov (BK) equation~\cite{Balitsky:1995ub,Kovchegov:1999yj}
\begin{eqnarray}
 {\ud\over \ud Y} S (\xt - \yt) =  
{\nc\, \as \over 2\pi^2} 
\int_\zt {\cal K}_{{\bf x} {\bf y} {\bf z}} 
\left[S (\xt - \zt)\, S (\zt - \yt) - S (\xt - \yt)\right].
\label{eq:2pt}
\end{eqnarray}
where ${\cal K}_{{\bf x} {\bf y} {\bf z}}=(\xt - \yt)^2 / (\xt - \zt)^2 (\zt - \yt)^2$.  A significant amount of work is underway in 
performing next-to-leading log $x$ extensions of the BK equation~\cite{NLL}-for the state of the art, we refer the reader to ref.~\cite{Balitsky:2012bs} and references therein. At the phenomenological level, a running coupling BK equation has been developed (rcBK)~\cite{Albacete:2007yr}, which gives good agreement with the latest combined H1 and ZEUS data sets~\cite{Aaron:2009aa}, and does better than NNLO DGLAP fits in the small $x$, small $Q^2$ region of the data~\cite{Albacete:2012rx}. 

For less inclusive observables, one encounters the expectation value $Q=\langle {\hat Q}\rangle$ of  the ``quadrupole" operator
\begin{equation}
{\hat Q}(\yt,\xt,\ut,\vt) = {1\over \nc} \tr \left(V^\dagger(\yt) V(\xt) V^\dagger (\ut) V (\vt)\right)  \, .
\label{eq:quad}
\end{equation}
Unlike $\langle {\hat S} {\hat S}\rangle$, it is not reducible to the product of dipoles even in the large $\nc$ and large $A$ approximations and is a novel universal correlator in high energy QCD~\cite{JalilianMarian:2004da,Dominguez:2011wm}, interesting both from theoretical and phenomenological perspectives. Examples of where this quantity appears include small-$x$ di-jet production in e+A DIS~\cite{Dominguez:2011wm}, quark-antiquark pair production in hadronic collisions~\cite{Blaizot:2004wv} and near-side long-range rapidity correlations~\cite{Dumitru:2010mv}. Another interesting quantity~\cite{Dominguez:2011wm} is a six-point correlator $S_6 \propto \langle {\hat Q} {\hat S}\rangle$ that is probed in forward di-hadron production $d+A\longrightarrow h_1\;h_2\;X$ in deuteron-gold collisions at RHIC.  The dominant underlying QCD process is the scattering of a large $x_1$ valence quark from the deuteron off small $x_2$ partons in the nuclear target, with the emission of a gluon from the valence quark either before or after the collision. The RHIC experiments show that the away-side peak in the di-hadron correlations is significantly broadened for central collisions at forward rapidities~\cite{dA-expt} as predicted in the CGC framework~\cite{Marquet:2007vb} and confirmed in more detailed analyses~\cite{dihadRHIC}. 

\begin{figure}[htb]
\begin{minipage}{0.5\linewidth}
\includegraphics[width=4cm, clip,trim=3.5cm 0cm 0cm 1cm]{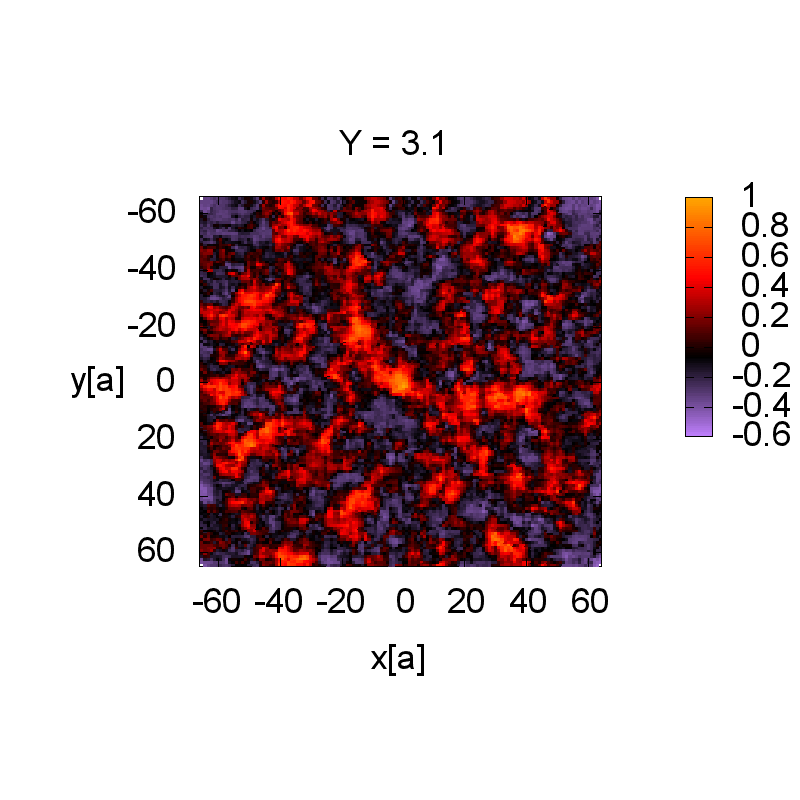}
\end{minipage}
\hskip 0.5in
\begin{minipage}{0.5\linewidth}
\includegraphics[width=70mm]{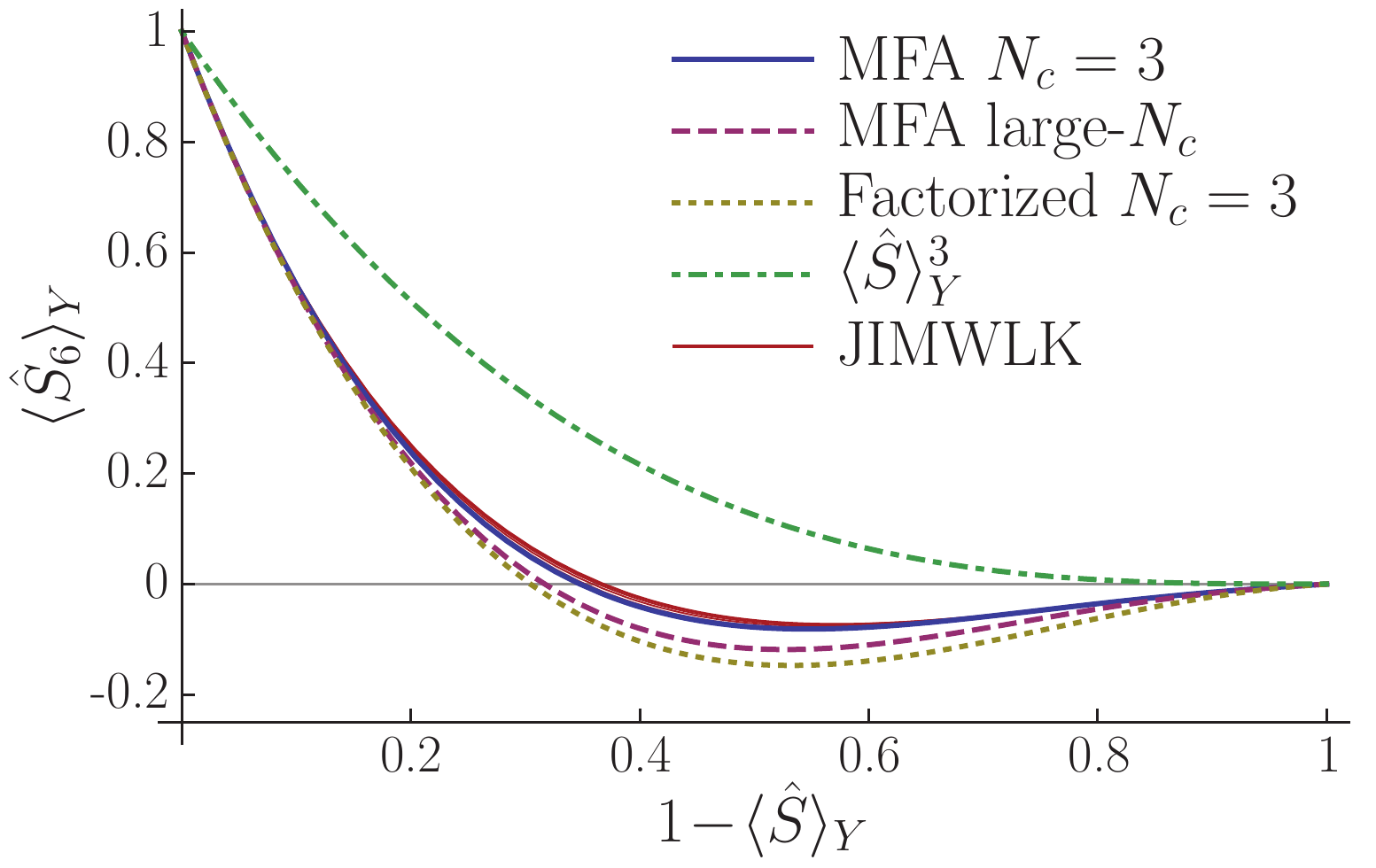}
\end{minipage}
\caption{Left: Correlation $1/\nc\langle V^\dag(0,0) V(x,y)\rangle$ between the center position $(0,0)$ and the point $(x,y)$ for a given rapidity $Y$. Colors illustrate the degree of fluctuations. 
From ref.~\cite{Dumitru:2011vk}. Right:  JIMWLK numerical simulation for  a particular ``line" configuration of the six point Wilson line correlator $\langle \hat{S}_6\rangle$ versus $1-\langle \hat{S}\rangle$, compared to an analytic Gaussian mean field approximation (MFA) for 
six different rapidities. From ref.~\cite{Iancu:2011nj}.}
\label{fig:JIMWLK}
\end{figure}

These analyses however relied on 
simplified factorization assumptions that are not justified. The JIMWLK RG equations for  $Q$~\cite{Dominguez:2011gc} and $S_6$~\cite{Dumitru:2010ak} have been derived and computed numerically for particular configurations of these operators~\cite{Dumitru:2011vk}. A noteworthy result of these numerical simulations is that the initial condition and the RG evolution of higher point correlators are well reproduced by expressions that are functions only of the dipole expectation value $S$. This result indicates strongly that the B-JIMWLK hierarchy is closely approximated by a Gaussian effective theory for $W[\rho]$ whose variance is proportional to $S$:  the energy evolution of the theory is controlled entirely by BK-evolution~\cite{Iancu:2011nj}. 
These results are illustrated in fig.~\ref{fig:JIMWLK}; the left figure shows the spatial (dipole) correlator of Wilson lines for a particular configuration of color charges from the numerical simulations of JIMWLK. The right part of the figure shows a comparison of the numerical results to the Gaussian approximation.  Excellent agreement is observed for the expectation value $S_6$ of the operator  appearing in di-hadron correlations, while the factorized expression $\langle \hat{S}\rangle^3$ disagrees strongly with the numerical results. 

The framework we have outlined here in terms of ``dipole" and ``quadrupole" operators has been used recently in phenomenological computations to describe the recent d+Au RHIC data~\cite{Stasto:2011ru}. More detailed computations and 
predictions for p+Pb results at the LHC are desirable~\cite{Kutak:2012rf}.

As this brief discussion suggests, the CGC approach to gluon saturation has been very fruitful with rapid ongoing developments in our understanding of the structure of matter at high energies. How does one best relate these developments to the outstanding issues raised at the outset ? There are many theoretical approaches feasible, but to my mind high luminosity studies of exclusive processes at a future Electron Ion Collider~\cite{EIC} offer the greatest promise of unraveling the spatial dependence of high energy evolution. This relates the weak coupling non-perturbative dynamics of gluon saturation 
to the intrinsically non-perturbative dynamics of chiral symmetry breaking and confinement, as exemplified in the high energy limit by the Froissart bound~\cite{Froissart:1961ux}.

We shall now discuss some of the implications of the the weak coupling ideas of the CGC EFT for high energy hadron scattering, in particular for nucleus-nucleus collisions.  In this framework, the collision can, to leading order,  be described as a collision of ``shock waves" of classical gluon fields~\cite{CYM}. The preequilibrium matter produced is called the Glasma~\cite{Glasma} and its properties can be determined {\it ab initio} in the CGC EFT framework. The classical field configurations are highly unstable; small quantum fluctuations around 
the classical fields are unstable and grow nearly exponentially~\cite{Weibel}. This rapid growth can cause the highly anisotropic initial Glasma color field configurations to isotropize on extremely rapid intervals not significantly greater than $1/Q_s$~\cite{Romatschke:2005pm}. This picture of underlying expanding classical fields and rapidly growing quantum fluctuations, whose interactions with the classical field isotropize the fields on very short time scales, bears strong analogy to ``pre-heating" inflationary dynamics in the very early universe that generates the ``hot" era of thermalized matter~\cite{inflation}. In particular, in direct analogy to the inflationary framework, preequilibrium dynamics is driven by a spectrum of amplified initial quantum fluctuations~\cite{Dusling:2010rm,Dusling:2012ig} potentially generating a significant amount of flow before thermalization.

 A master formula which incorporates the leading divergent contributions from quantum fluctuations, both before and after the collision, has been derived to leading accuracy in a weak coupling expansion of inclusive quantitities; for the components of the stress-energy tensor, one obtains~\cite{Dusling:2011rz}
\begin{equation}
\langle T^{\mu\nu}\rangle_{{\rm LLx + LInst.}} =\int [D\rho_1 D\rho_2]\;
W_{x_1}[\rho_1]\, W_{x_2}[\rho_2]  \int \!\! \big[{\cal D}\alpha\big]\,
F_0\big[\alpha\big]\; T_{_{\rm LO}}^{\mu\nu} [{\cal A}[\rho_1,\rho_2] + \alpha] (x)\; .
\label{eq:final-formula}
\end{equation}
The argument ${\cal A}\equiv ( A, E)$ denotes collectively the components of the classical fields and their canonically
conjugate momenta on the initial proper time surface; analytical expressions for these are available at
$\tau=0^+$~\cite{CYM}. Their temporal evolution is obtained by solving  Yang-Mills equations with static light front color  sources~\cite{Yang-Mills}. The $W$'s are the functional density matrices we defined previously that obey the JIMWLK equation. To leading log accuracy in $x$, the $W$'s for the two nuclei can be factorized~\cite{Factorization}. The initial spectrum of fluctuations $F_0\big[\alpha\big]$, Gaussian in the quantum fluctuations $\alpha$,  has a variance given by the small fluctuation propagator in the Glasma background field as $\tau\rightarrow 0^+$. In practice, the path integral in $\alpha$ is determined by solving the classical Yang-Mills equations repeatedly with the initial conditions at $\tau=0^+$ given by
\begin{equation}
{\bf A}_{\rm init.} = {\cal A}_{\rm init.} + \int d\mu_{_K}\;\Big[c_{_K}\,a_{_K}^\mu(x)+c_{_K}^*\,a_{_K}^{\mu*}(x)\Big] \, .
\label{eq:quantum}
\end{equation}
Here ${\bf A}$ collectively denotes the quantum fields and their canonical conjugate momenta. The  coefficients $c_{_K}$, with $K$ collectively denoting the quantum numbers labeling the basis of solutions, are random Gaussian-distributed complex numbers. Explicit expressions for the small fluctuations and their conjugate momenta, denoted here by $a_K^\mu (x)$ were obtained in ref.~\cite{Dusling:2011rz}. 

\begin{figure}[htb]
\centering
\includegraphics[width=65mm]{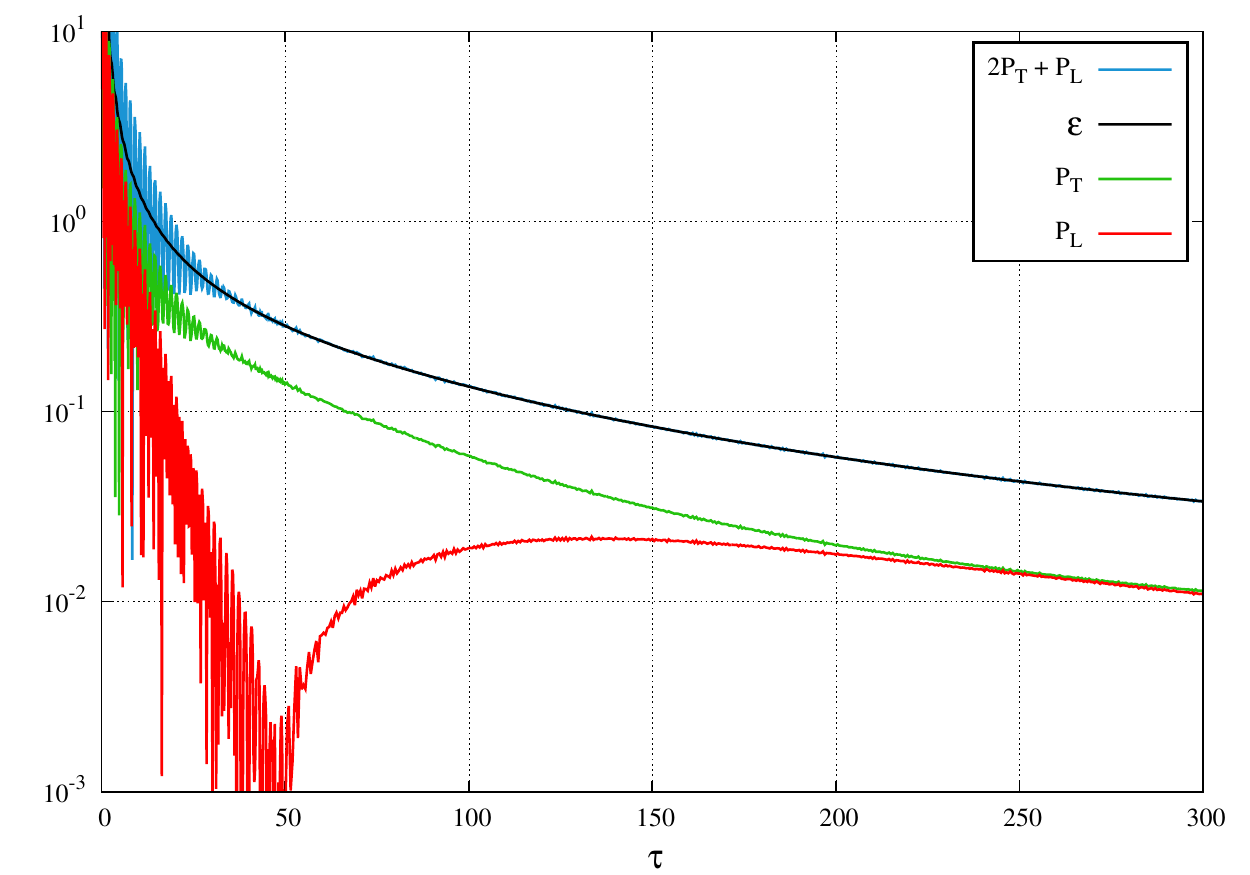}
\hskip 0.5in
\includegraphics[width=70mm]{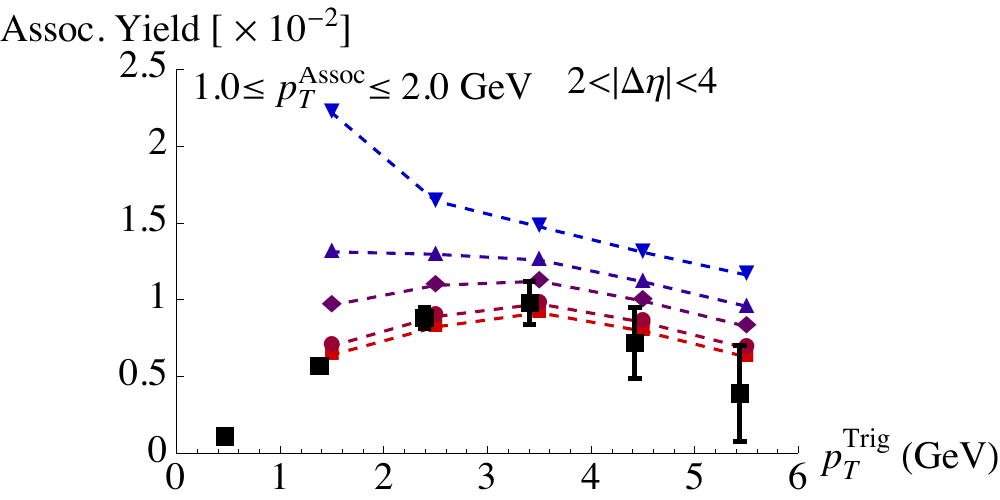}
\caption{Left: Time evolution of the diagonal elements of $T^{\mu\nu}$ and of the trace of the pressure tensor for a longitudinally expanding $\Phi^4$ theory. Figure from ref.~\cite{Dusling:2012-flow}. Right: Nearside di-hadron yield in high multiplicity collisions from Glasma graphs. The effect of radial flow on this contribution is seen with increasing flow velocity $\beta$ (from bottom to top)=
0.0,0.1,0.2,0.25,0.3. Figure from ref.~\cite{Dusling:2012ig}.
}
\label{fig:tubes}
\end{figure}

The averaging over multiple initializations of the field leads to decoherence of the classical fields, coinciding with the development of an equation of state for the (classicaly) conformal field theories considered and eventually isotropization and thermalization of the system. In ref.~\cite{Dusling:2012-flow}, based on the formalism outlined here, it was shown explicitly that a longitudinally expanding scalar $\phi^4$ theory isotropizes despite a very rapid initial red shift of the longitudinal pressure.  This result is shown in fig.~\ref{fig:tubes} (left), where the time scale is in lattice units. This gives us hope that the same will occur for the gauge theory case as well where numerical work is in progress. 

An outstanding problem is how the system evolving with the above initial conditions will thermalize. Clearly this has important ramifications for when hydrodynamics is applicable as a description of heavy ion collisions. Recently there has been significant progress in understanding how thermalization is achieved for a fixed 
box both from transport computations~\cite{transport} and classical lattice computations~\cite{SFT}. The situation for a longitudinally expanding gauge theory is still fluid with a clear resolution likely in the near future~\cite{Berges:2012ks}. 

An important feature of the Glasma is that it generates long-range correlations that are localized on scales $\sim 1/Q_S$ in the transverse plane~\cite{Dumitru:2008wn}. Based on the formalism in ref.~\cite{Factorization}, its feasible to study the evolution of these correlations with energy and the the rapidity separation between correlated gluons~\cite{Dusling:2009ni}. Two particle correlations that are azimuthally collimated around $\Delta \phi \sim 0$ and are long range in their rapidity separation $\Delta \eta$ were observed in high multiplicity p+p collisions by the CMS collaboration~\cite{CMS-ridge}. Though they appear on the surface similar to phenomena observed in A+A collisions at RHIC~\cite{RHIC-ridge}, their origin is qualitatively different. 
As argued in \cite{Dumitru-etal}, the effect is due to doubly inclusive $1/\nc^2$ suppressed ``Glasma" graphs that are enhanced strikingly by $\alpha_S^{-8}$ (a factor of $10^4$ -- $10^5$ for a reasonable range in the coupling) in high multiplicity events for momenta $\leq Q_S$. This enhancement corresponds to effective strong  color source densities being probed of order $\sim 1/\sqrt{\alpha_S}$ in the colliding protons. This explanation was shown to be quantitatively correct in ref.~\cite{Dusling:2012ig} from a comparison to the systematics of the per trigger near side yield. It was also shown in this work that increasing radial flow changes the behavior of the associated yield as a function of trigger particle $p_T^{\rm trig}$, in disagreement with data. This result is shown in fig.~\ref{fig:tubes} (right). 
In contrast, the same analysis for the CMS associated yield, in the same kinematic window, for A+A collisions shows that the magnitude and shape of the yield with $p_T^{\rm trig.}$ agrees with a flow analysis requiring a large collective flow velocity. To summarize, the high multiplicity $p+p$ ridge exhibits an azimuthal near side collimation due to a subtle non-trivial manifestation of saturation dynamics in multi-particle production. The $A+A$ ridge, while fundamentally also due to the Glasma dynamics generating long range correlations, exhibits (a much larger) azimuthal collimation primarily due to collective radial flow of hot matter. 

\begin{figure}[htb]
\begin{minipage}{0.5\linewidth}
\centering
\includegraphics[width=4cm, clip,trim=3.5cm 0cm 0cm 1cm]{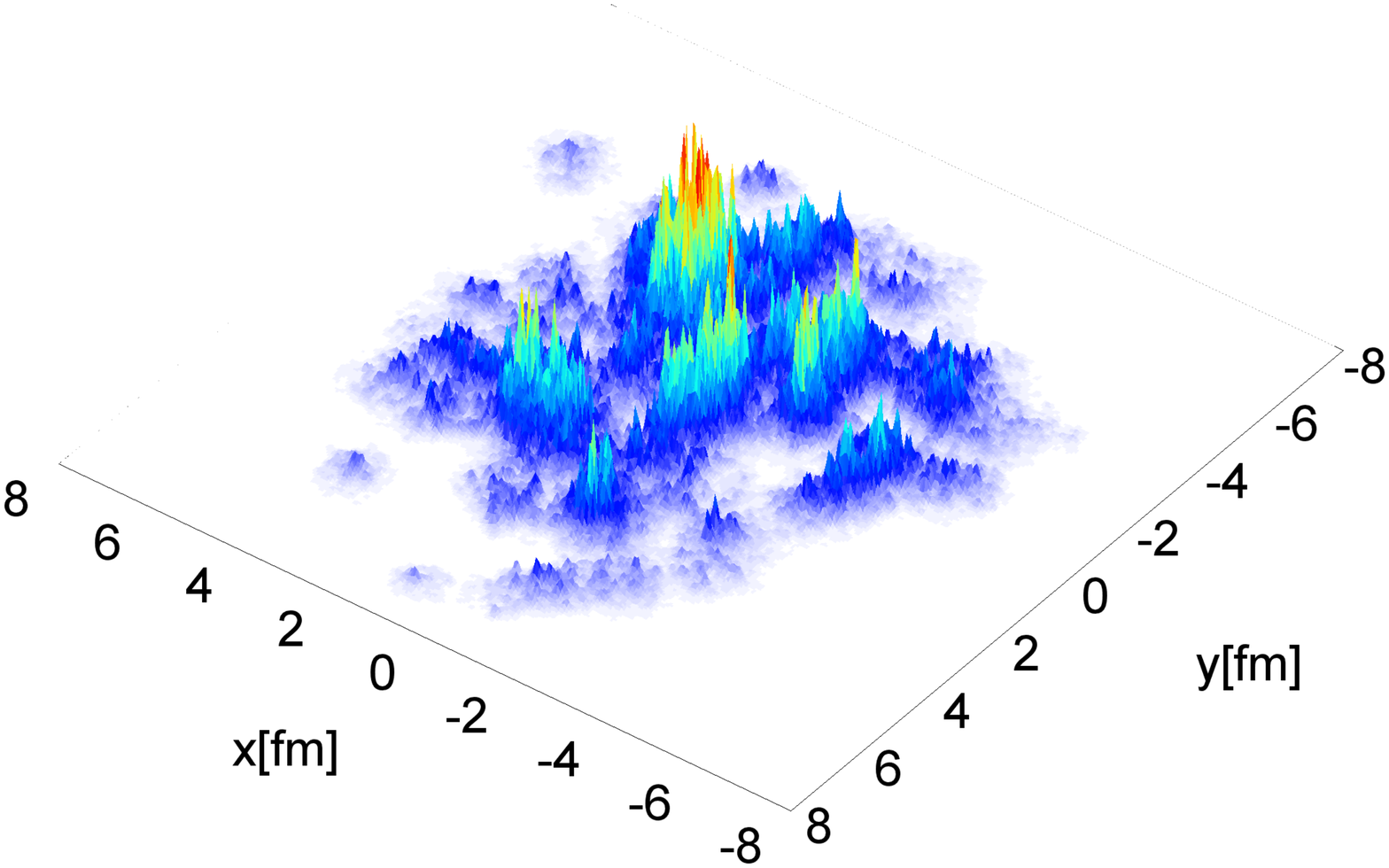}
\vspace{-0.9in}
\includegraphics[width=4cm, clip,trim=3.5cm 0cm 0cm 1cm]{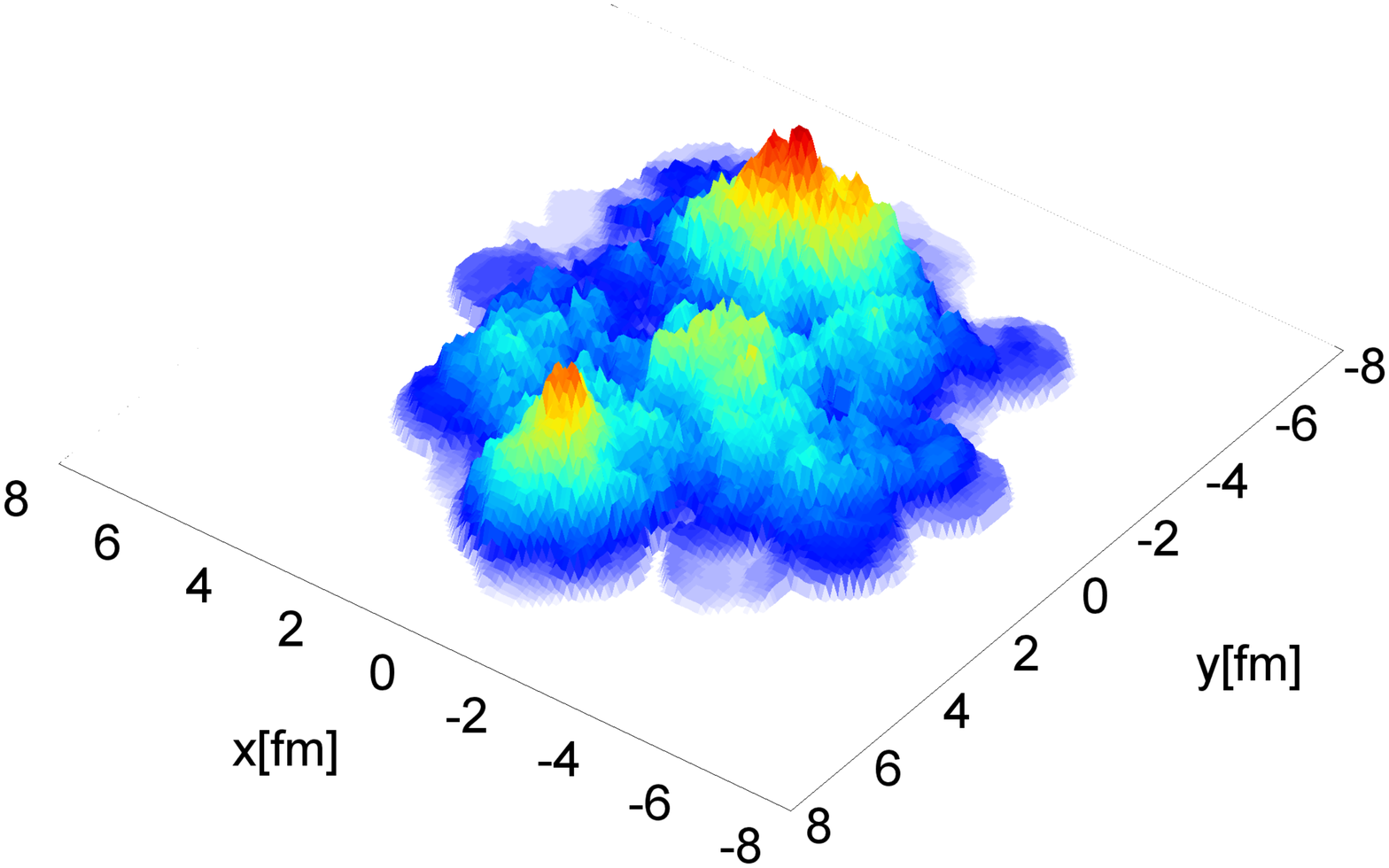}
\end{minipage}
\hskip 0.5in
\begin{minipage}{0.5\linewidth}
\includegraphics[width=70mm]{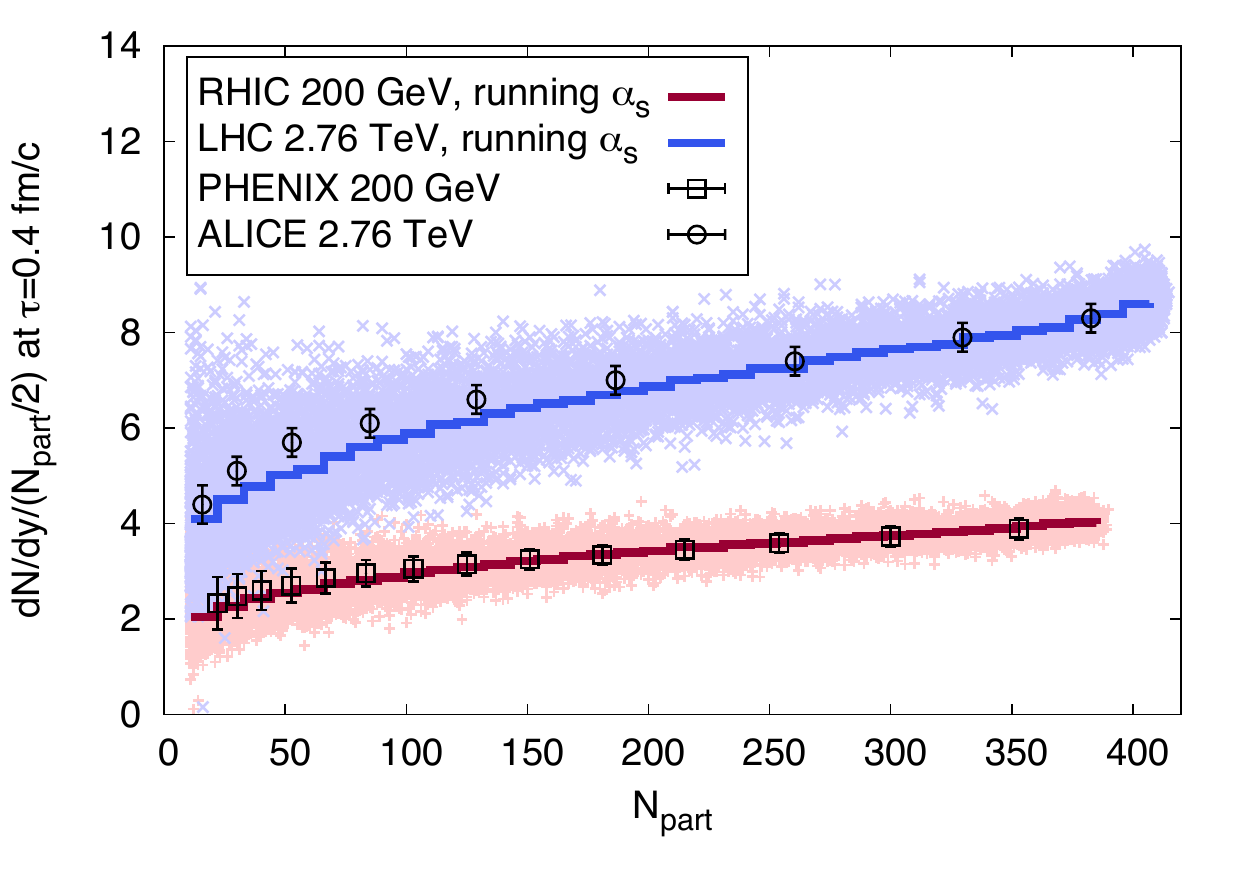}
\end{minipage}
\vspace{0.5in}
\caption{Left: Initial energy density distribution in three different heavy ion collision events. Top:IP-Glasma model. Bottom:MC-KLN model. Figure from ref.~\cite{Schenke:2012wb}.  Right: Gluon multiplicity $(dN_g/dy)/(N_{\rm part}/2)$ at $\tau=0.4\,{\rm fm}/c$ times $2/3$ compared to experimental charged particle $(dN/dy)/(N_{\rm part}/2)$ data for $\sqrt{s}=200\,{\rm GeV}$ Au+Au
    and $\sqrt{s}=2.76\,{\rm TeV}$ Pb+Pb collisions as a function of $N_{\rm part}$. The pale blue and red bands are a collection of the multiplicities for individual events, with the solid lines representing the average multiplicity. Figure from ref.~\cite{Schenke:2012hg}.
}
\label{fig:IPglasma}
\end{figure}

As our last topic, we will discuss a phenomenological model of how strongly correlated gluodynamics in nuclear wavefunctions generates, event-by-event, entropy production and flow in heavy ion collisions across a range of centralities and energies.
This model, called the IP-Glasma model, incorporates a number of sources of event-by-event fluctuations in multiparticle production. A significant source of fluctuations are those of the nucleon positions from event-to-event. In addition, there are event-by-event fluctuations in the color charge distributions in each of the nucleons. Finally, there are fluctuations in the number of gluons produced from event-to-event. The color charge fluctuations in the IP-Glasma model are 
constrained by the IP-Sat model (which incorporates energy and impact parameter dependence in the McLerran-Venugopalan  model) ~\cite{IPsat} fits to HERA inclusive and diffractive DIS data on e+p collisions, and the limited fixed target e+A DIS data available. Convoluting these {\it nucleon} color charge fluctuations with Woods-Saxon distributions of nucleons, one constructs Lorentz contracted two dimensional nuclear color charge distributions of the incoming nuclei event-by-event. In contrast to prior models, the scale of transverse event-by-event fluctuations is $1/Q_S$  and not the nucleon size.
This IP-Glasma model~\cite{Schenke:2012wb,Schenke:2012hg} employs the fluctuating gluon fields generated by the IP-Sat model to study the event-by-event 2+1-D Yang-Mills  evolution of gluon fields in the Glasma. The corresponding energy density distributions in the Glasma vary on the scale $1/Q_S$ and are therefore much more highly localized than in models that do not include color charge fluctuations. For a comparison of the initial energy density distribution in the IP-Glasma model to the CGC motivated MC-KLN model~\cite{mckln} that does not include color charge fluctuations and multiplicity fluctuations on size scales $1/Q_S$, see fig.~\ref{fig:IPglasma} (left). 

The IP-Glasma model does an excellent job of describing the single inclusive multiplicity as a function of centrality at both RHIC and the LHC. We find that running coupling effects are important for a good description of data. (See fig.~\ref{fig:IPglasma} (right).) We also find that the event-by-event IP Glasma does an excellent job of describing the n-particle multiplicity distributions measured at RHIC and the LHC. 
In ref.~\cite{Gelis:2009wh}, it was shown in a perturbative computation that the n-particle distribution of the ``Glasma flux tube" 
framework of ref.~\cite{Dumitru:2008wn} is a negative binomial distribution (NBD), where the variables are the average multiplicity and the parameter $k=\zeta (\nc^2 -1) Q_S^2\, S_\perp/2\pi$, with $S_\perp$ the transverse overlap area in the collision and $\zeta$ a non-perturbative parameter to be determined numerically. A Yang-Mills computation of the double inclusive gluon distribution~\cite{Lappi:2009xa} shows that this result is non-perturbatively robust with $\zeta\sim 0.3$-$1.5$. The $n$-particle distribution is extracted numerically in ref.~\cite{Schenke:2012wb} from event by event solutions of Yang-Mills equations. The distribution obeys the NBD form, with $\zeta\sim 0.2$ for smooth distributions at large values of $Q_S^2 \, S_\perp$. These descriptions of the energy and centrality dependence of multiplicity distributions (for further examples, see refs.~\cite{LHC-pheno}) are strong indications that the CGC provides the right framework for entropy production. 

Turning to flow, a sensitive measure is provided by harmonic flow coefficients $v_n$ of the anisotropic azimuthal single particle distribution that are related to the initial spatial eccentricities $\varepsilon_n$ by hydrodynamic flow. How efficiently this is done is a measure of the transport properties of strongly correlated QCD matter such as the shear and bulk viscosity to entropy density ratios~\cite{Schenke:2011qd} . In ref.~\cite{Schenke:2012hg}, results were presented for $\varepsilon_2$ through $\varepsilon_6$. The conclusions that can be drawn are mainly that the purely fluctuation driven odd harmonics $\varepsilon_3$ and $\varepsilon_5$ from the IP-Glasma model are larger than those from the MC-KLN model. In contrast,  $\varepsilon_2$ is smaller than that computed in the MC-KLN model, in particular for larger impact parameters.
As a consequence, the ratio $\varepsilon_2/\varepsilon_3$ is smaller than in the MC-KLN model, which decreases the ratio of $v_2/v_3$ obtained after hydrodynamic evolution, thereby making it more compatible with experimental observations.

\hspace{-0.65cm}{\it \bf Acknowledgments:} I thank my collaborators J.-P. Blaizot, A. Dumitru, K. Dusling, T. Epelbaum, F. Gelis, J.Jalilian-Marian, T. Lappi, J. Liao, L. McLerran, B. Schenke, S. Schlichting and P. Tribedy.  I would also like to thank the organizers, in particular Bernard Pire and Michel Gar\c{c}on, for their kind invitation to speak at this interesting conference covering a wide range of topics in hadronic physics.

\end{document}